\documentclass[12pt,reqno]{amsart}
\usepackage{amsfonts}
\usepackage{amssymb,amsmath}
\usepackage{amscd}
\usepackage{hyperref}
\usepackage{graphicx}
\usepackage{epsfig}

\newtheorem{example}{Example}[section]

\numberwithin{equation}{section}
\begin{document}
\title[A New Cryptography Model via Fibonacci and Lucas Numbers]{A New
Cryptography Model via Fibonacci and Lucas Numbers}
\author[S. U\c{C}AR]{S\"{U}MEYRA U\c{C}AR*}
\address{Bal\i kesir University\\
Department of Mathematics\\
10145 Bal\i kesir, TURKEY}
\email{sumeyraucar@balikesir.edu.tr}
\thanks{*Corresponding author: S. U\c{C}AR\\
Bal\i kesir University, Department of Mathematics, 10145 Bal\i kesir, TURKEY%
\\
e-mail: sumeyraucar@balikesir.edu.tr}
\author[N. TA\c{S}]{N\.{I}HAL TA\c{S}}
\address{Bal\i kesir University\\
Department of Mathematics\\
10145 Bal\i kesir, TURKEY}
\email{ nihaltas@balikesir.edu.tr}
\author[N. YILMAZ \"{O}ZG\"{U}R]{N\.{I}HAL YILMAZ \"{O}ZG\"{U}R}
\address{Bal\i kesir University\\
Department of Mathematics\\
10145 Bal\i kesir, TURKEY}
\email{nihal@balikesir.edu.tr}
\date{}

\begin{abstract}
Coding/decoding algorithms are of great importance to help in improving
information security since information security is a more significiant
problem in recent years. In this paper we introduce two new coding/decoding
algorithms using Fibonacci $Q$-matrices and $R$-matrices. Our models are
based on the blocked message matrices and the encryption of each message
matrix with different keys. These new algorithms will not only increase the
security of information but also has high correct ability.
\end{abstract}

\maketitle

\section{Introduction}

\label{intro}

It is well known that the sequences of Fibonacci and Lucas numbers are
defined by 
\begin{equation}
F_{n+1}=F_{n}+F_{n-1}\text{,}  \label{eqn1}
\end{equation}%
\begin{equation*}
L_{n+1}=L_{n}+L_{n-1}
\end{equation*}%
with the initial terms $F_{0}=0$, $F_{1}=1$ and $L_{0}=2$, $L_{1}=1$,
respectively (see \cite{koshy} for more details). The Fibonacci $Q$-matrix
is defined in \cite{gould} and \cite{hoggat} as follows:%
\begin{equation*}
Q=\left[ 
\begin{array}{cc}
1 & 1 \\ 
1 & 0%
\end{array}%
\right] .
\end{equation*}

From \cite{stakhov 1999} and \cite{stakhov 2006}, we known that the $n.$th
power of the Fibonacci $Q$-matrix is of the following form:%
\begin{equation*}
Q^{n}=\left[ 
\begin{array}{cc}
F_{n+1} & F_{n} \\ 
F_{n} & F_{n-1}%
\end{array}%
\right] \text{.}
\end{equation*}

In \cite{brugless}, Buggles and Hoggat introduced the $R$-matrix as follows$%
: $%
\begin{equation*}
R=\left[ 
\begin{array}{cc}
1 & 2 \\ 
2 & -1%
\end{array}%
\right]
\end{equation*}%
Using the Fibonacci $Q$-matrix and $R$-matrix, it was obtained the matrix $%
R_{n}$ of the following form$:$ 
\begin{equation*}
R_{n}=RQ^{n}=\left[ 
\begin{array}{cc}
1 & 2 \\ 
2 & -1%
\end{array}%
\right] \left[ 
\begin{array}{cc}
F_{n+1} & F_{n} \\ 
F_{n} & F_{n-1}%
\end{array}%
\right] =\left[ 
\begin{array}{cc}
L_{n+1} & L_{n} \\ 
L_{n} & L_{n-1}%
\end{array}%
\right] \text{.}
\end{equation*}

Determinants of the Fibonacci $Q$-matrix and the $R$-matrix are as follows:%
\begin{equation*}
Det(Q^{n})=F_{n+1}F_{n-1}-F_{n}^{2}=(-1)^{n}
\end{equation*}%
and%
\begin{equation*}
Det(R_{n})=L_{n+1}L_{n-1}-L_{n}^{2}=5(-1)^{n+1}\text{.}
\end{equation*}

Fibonacci coding theory was studied by different ways. For example, in \cite%
{prajat} a new approach for secure information transmission over
communication channel was obtained with key variability concept in symmetric
key algorithms using Fibonacci $Q$-matrix. In \cite{stakhov 2006}, a new
coding theory was introduced using the generalization of the Cassini formula
for Fibonacci $p$-numbers and $Q_{p}$-matrices. In \cite{Wang}, it was
constructed an application of mobile phone encryption based on Fibonacci
structure of chaos using Fibonacci series. In \cite{prasad-lucas}, Prasad
developed a new coding and decoding method using Lucas $p$ numbers given in 
\cite{Kuhapatanakul}. Recently, a new cryptography algorithm has been
introduced by blocking matrices and Fibonacci numbers in \cite{Tas}. Also
there are more studies in the literature (see \cite{basu}, \cite%
{stakhov1999-2}, \cite{Tarle} and the references therein for more details).

In this study we introduce two new coding/decoding algorithms using
Fibonacci $Q$-matrices and $R$-matrices. The basic idea of our method
depends on dividing the message matrix into the block matrices of size $%
2\times 2$. Because of using mixed type algorithm and different numbered
alphabet for each message, we have a more safely coding/decoding method. The
alphabet is determined by the number of block matrices of the message
matrix. Our method will not only increase the security of information but
also has high correct ability for data transfer over communication channel.

\section{A New Coding/Decoding Method using $R$-Matrix}

\label{sec:1}

In this section we introduce a new coding/decoding algorithm using Lucas
numbers. We put our message in a matrix of even size adding zero between two
words and end of the message until we obtain the size of the message matrix
is even. Dividing the message square matrix $M$ of size $2m$ into the block
matrices, named $B_{i}$ ($1\leq i\leq m^{2}$) of size $2\times 2$, from left
to right, we construct a new coding method.

Now we explain the symbols of our coding method. Assume that matrices $B_{i}$%
, $E_{i}$, $Q^{n}$ and $R_{n}$ are of the following forms:%
\begin{equation*}
B_{i}=\left[ 
\begin{array}{cc}
b_{1}^{i} & b_{2}^{i} \\ 
b_{3}^{i} & b_{4}^{i}%
\end{array}%
\right] \text{, }E_{i}=\left[ 
\begin{array}{cc}
e_{1}^{i} & e_{2}^{i} \\ 
e_{3}^{i} & e_{4}^{i}%
\end{array}%
\right] \text{, }Q^{n}=\left[ 
\begin{array}{cc}
q_{1} & q_{2} \\ 
q_{3} & q_{4}%
\end{array}%
\right] \text{ and }R_{n}=\left[ 
\begin{array}{cc}
r_{1} & r_{2} \\ 
r_{3} & r_{4}%
\end{array}%
\right] \text{.}
\end{equation*}%
The number of the block matrices $B_{i}$ is denoted by $b$. According to $b$%
, we choose the number $n$ as follows:%
\begin{equation*}
n=\left\{ 
\begin{array}{ccc}
b & \text{,} & b\leq 3 \\ 
\left[ \left\vert \frac{b}{2}\right\vert \right] & \text{,} & b>3%
\end{array}%
\right. \text{.}
\end{equation*}%
Using the chosen $n$, we write the following character table according to $%
mod30$ (this table can be extended according to the used characters in the
message matrix). We begin the \textquotedblleft $n$\textquotedblright\ for
the first character.

\begin{equation*}
\begin{tabular}{|c|c|c|c|c|c|c|c|c|c|}
\hline
A & B & C & D & E & F & G & H & I & J \\ \hline
$n$ & $n+1$ & $n+2$ & $n+3$ & $n+4$ & $n+5$ & $n+6$ & $n+7$ & $n+8$ & $n+9$
\\ \hline
K & L & M & N & O & P & Q & R & S & T \\ \hline
$n+10$ & $n+11$ & $n+12$ & $n+13$ & $n+14$ & $n+15$ & $n+16$ & $n+17$ & $%
n+18 $ & $n+19$ \\ \hline
U & V & W & X & Y & Z & 0 & ! & ? & . \\ \hline
$n+20$ & $n+21$ & $n+22$ & $n+23$ & $n+24$ & $n+25$ & $n+26$ & $n+27$ & $%
n+28 $ & $n+29$ \\ \hline
\end{tabular}%
\end{equation*}

Now we explain the following new coding and decoding algorithms.

\textbf{Lucas Blocking Algorithm}

\textbf{Coding Algorithm}

\textbf{Step 1.} Divide the matrix $M$ into blocks $B_{i}$ $\left( 1\leq
i\leq m^{2}\right) $.

\textbf{Step 2.} Choose $n$.

\textbf{Step 3. }Determine $b_{j}^{i}$ $\left( 1\leq j\leq 4\right) $.

\textbf{Step 4.} Compute $\det (B_{i})\rightarrow d_{i}$.

\textbf{Step 5.} Construct $F=\left[ d_{i},b_{k}^{i}\right] _{k\in
\{1,2,4\}} $.

\textbf{Step 6. }End of algorithm.

\textbf{Decoding Algorithm }

\textbf{Step 1.} Compute $R_{n}$.

\textbf{Step 2.} Determine $r_{j}$ $(1\leq j\leq 4)$.

\textbf{Step 3. }Compute $r_{1}b_{1}^{i}+r_{3}b_{2}^{i}\rightarrow e_{1}^{i}$
$\left( 1\leq i\leq m^{2}\right) $.

\textbf{Step 4.} Compute $r_{2}b_{1}^{i}+r_{4}b_{2}^{i}\rightarrow e_{2}^{i}$%
.

\textbf{Step 5.} Solve $5\times (-1)^{n+1}\times
d_{i}=e_{1}^{i}(r_{2}x_{i}+r_{4}b_{4}^{i})-e_{2}^{i}(r_{1}x_{i}+r_{3}b_{4}^{i}) 
$.

\textbf{Step 6. }Substitute for $x_{i}=b_{3}^{i}$.

\textbf{Step 7.} Construct $B_{i}$.

\textbf{Step 8.} Construct $M$.

\textbf{Step 9.} End of algorithm.

The above method is similar to the method obtained by Fibonacci numbers
given in \cite{Tas}.

In the following example we give an application of the above algorithm for $%
b>3$.

\begin{example}
\label{exm1} Let us consider the message matrix for the following message
text$:$%
\begin{equation*}
\text{\textquotedblleft HI! HOW ARE YOU?\textquotedblright }
\end{equation*}%
Using the message text, we get the following message matrix $M:$%
\begin{equation*}
M=\left[ 
\begin{array}{cccc}
H & I & ! & 0 \\ 
H & O & W & 0 \\ 
A & R & E & 0 \\ 
Y & O & U & ?%
\end{array}%
\right] _{4\times 4}.
\end{equation*}%
\textbf{Coding Algorithm:}

\textbf{Step 1. }We can divide the message matrix $M$ of size $4\times 4$
into the matrices, named $B_{i}$ $\left( 1\leq i\leq 4\right) $, from left
to right, each of size is $2\times 2:$%
\begin{equation*}
B_{1}=\left[ 
\begin{array}{cc}
H & I \\ 
H & O%
\end{array}%
\right] \text{, }B_{2}=\left[ 
\begin{array}{cc}
! & 0 \\ 
W & 0%
\end{array}%
\right] \text{, }B_{3}=\left[ 
\begin{array}{cc}
A & R \\ 
Y & O%
\end{array}%
\right] \text{ and }B_{4}=\left[ 
\begin{array}{cc}
E & 0 \\ 
U & ?%
\end{array}%
\right] \text{.}
\end{equation*}

\textbf{Step 2.} Since $b=4\geq 3$, we calculate $n=\left[ \left\vert \frac{b%
}{2}\right\vert \right] =2$. For $n=2$, we use the following
\textquotedblleft character table\textquotedblright\ for the message matrix $%
M:$%
\begin{equation*}
\begin{tabular}{|l|l|l|l|l|l|l|l|}
\hline
$H$ & $I$ & $!$ & $0$ & $H$ & $O$ & $W$ & $0$ \\ \hline
$9$ & $10$ & $29$ & $28$ & $9$ & $16$ & $24$ & $28$ \\ \hline
$A$ & $R$ & $E$ & $0$ & $Y$ & $O$ & $U$ & $?$ \\ \hline
$2$ & $19$ & $6$ & $28$ & $26$ & $16$ & $22$ & $0$ \\ \hline
\end{tabular}%
\text{.}
\end{equation*}

\textbf{Step 3.} We have the elements of the blocks $B_{i}$ $\left( 1\leq
i\leq 4\right) $ as follows:%
\begin{equation*}
\begin{tabular}{|l|l|l|l|}
\hline
$b_{1}^{1}=9$ & $b_{2}^{1}=10$ & $b_{3}^{1}=9$ & $b_{4}^{1}=16$ \\ \hline
$b_{1}^{2}=29$ & $b_{2}^{2}=28$ & $b_{3}^{2}=24$ & $b_{4}^{2}=28$ \\ \hline
$b_{1}^{3}=2$ & $b_{2}^{3}=19$ & $b_{3}^{3}=26$ & $b_{4}^{3}=16$ \\ \hline
$b_{1}^{4}=6$ & $b_{2}^{4}=28$ & $b_{3}^{4}=22$ & $b_{4}^{4}=0$ \\ \hline
\end{tabular}%
.
\end{equation*}

\textbf{Step 4.} Now we calculate the determinants $d_{i}$ of the blocks $%
B_{i}:$%
\begin{equation*}
\begin{tabular}{|l|}
\hline
$d_{1}=\det (B_{1})=54$ \\ \hline
$d_{2}=\det (B_{2})=140$ \\ \hline
$d_{3}=\det (B_{3})=-462$ \\ \hline
$d_{4}=\det (B_{4})=-616$ \\ \hline
\end{tabular}%
.
\end{equation*}

\textbf{Step 5.} Using Step 3 and Step 4 we obtain the following matrix $F:$%
\begin{equation*}
F=\left[ 
\begin{array}{cccc}
54 & 9 & 10 & 16 \\ 
140 & 29 & 28 & 28 \\ 
-462 & 2 & 19 & 16 \\ 
-616 & 6 & 28 & 0%
\end{array}%
\right] .
\end{equation*}

\textbf{Step 6.} End of algorithm.

\textbf{Decoding algorithm:}

\textbf{Step 1.} It is known that%
\begin{equation*}
R_{2}=RQ^{2}=\left[ 
\begin{array}{cc}
4 & 3 \\ 
3 & 1%
\end{array}%
\right] \text{.}
\end{equation*}

\textbf{Step 2. }The elements of $R_{2}$ are denoted by%
\begin{equation*}
r_{1}=4\text{, }r_{2}=3\text{, }r_{3}=3\text{ and }r_{4}=1\text{.}
\end{equation*}

\textbf{Step 3.} We compute the elements $e_{1}^{i}$ to construct the matrix 
$E_{i}:$%
\begin{equation*}
e_{1}^{1}=66\text{, }e_{1}^{2}=200\text{, }e_{1}^{3}=65\text{ and }%
e_{1}^{4}=108\text{.}
\end{equation*}

\textbf{Step 4. }We compute the elements $e_{2}^{i}$ to construct the matrix 
$E_{i}:$%
\begin{equation*}
e_{2}^{1}=37\text{, }e_{2}^{2}=115\text{, }e_{2}^{3}=25\text{ and }%
e_{2}^{4}=46\text{.}
\end{equation*}

\textbf{Step 5.} We calculate the elements $x_{i}:$%
\begin{eqnarray*}
5(-1)^{3}(54) &=&66(3x_{1}+16)-37(4x_{1}+48) \\
&\Rightarrow &x_{1}=9\text{.}
\end{eqnarray*}%
\begin{eqnarray*}
5(-1)^{3}(140) &=&200(3x_{2}+28)-115(4x_{2}+84) \\
&\Rightarrow &x_{2}=24\text{.}
\end{eqnarray*}%
\begin{eqnarray*}
5(-1)^{3}(-462) &=&65(3x_{3}+16)-25(4x_{3}+48) \\
&\Rightarrow &x_{3}=26\text{.}
\end{eqnarray*}%
\begin{eqnarray*}
5(-1)^{4}(-616) &=&108(3x_{4}+0)-46(4x_{4}+0) \\
&\Rightarrow &x_{4}=22\text{.}
\end{eqnarray*}

\textbf{Step 6.} We rename $x_{i}$ as follows$:$%
\begin{equation*}
x_{1}=b_{3}^{1}=9\text{, }x_{2}=b_{3}^{2}=24\text{, }x_{3}=b_{3}^{3}=26\text{
and }x_{4}=b_{3}^{4}=22\text{.}
\end{equation*}

\textbf{Step 7. }We construct the block matrices $B_{i}:$%
\begin{equation*}
B_{1}=\left[ 
\begin{array}{cc}
9 & 10 \\ 
9 & 16%
\end{array}%
\right] \text{, }B_{2}=\left[ 
\begin{array}{cc}
29 & 28 \\ 
24 & 28%
\end{array}%
\right] \text{, }B_{3}=\left[ 
\begin{array}{cc}
2 & 19 \\ 
26 & 16%
\end{array}%
\right] \text{ and }B_{4}=\left[ 
\begin{array}{cc}
6 & 28 \\ 
22 & 0%
\end{array}%
\right] \text{.}
\end{equation*}

\textbf{Step 8.} We obtain the message matrix $M:$%
\begin{equation*}
M=\left[ 
\begin{array}{cccc}
9 & 10 & 29 & 28 \\ 
9 & 16 & 24 & 28 \\ 
2 & 19 & 6 & 28 \\ 
26 & 16 & 22 & 0%
\end{array}%
\right] =\left[ 
\begin{array}{cccc}
H & I & ! & 0 \\ 
H & O & W & 0 \\ 
A & R & E & 0 \\ 
Y & O & U & ?%
\end{array}%
\right] .
\end{equation*}

\textbf{Step 9.} End of algorithm.
\end{example}

\section{A Mixed Model: Minesweeper Model}

\label{sec:2} In this section we present a new approach to coding/decoding
algorithm method called as \textquotedblleft Minesweeper
Model\textquotedblright\ using Fibonacci $Q^{n}$-matrices and $R$-matrices.
The main idea of this model is to decode blocks of the message matrix using
Fibonacci and Lucas numbers randomly. In the following model is constructed
by decoding the blocks with odd indices $i$ using Fibonacci $Q^{n}$-matrices
and decoding the blocks with even indices $i$ using $R$-matrices.

\textbf{Minesweeper Algorithm}

\textbf{Coding Algorithm}

\textbf{Step 1.} Divide the matrix $M$ into blocks $B_{i}$ $\left( 1\leq
i\leq m^{2}\right) $.

\textbf{Step 2.} Choose $n$.

\textbf{Step 3. }Determine $b_{j}^{i}$ $\left( 1\leq j\leq 4\right) $.

\textbf{Step 4.} Compute $\det (B_{i})\rightarrow d_{i}$.

\textbf{Step 5.} Construct $F=\left[ d_{i},b_{k}^{i}\right] _{k\in
\{1,2,3\}} $.

\textbf{Step 6. }End of algorithm.

\textbf{Decoding Algorithm }

\textbf{Step 1.} Compute $Q^{n}$.

\textbf{Step 2. }Compute $R^{n}$.\textbf{\ }

\textbf{Step 3. }Compute $q_{1}b_{1}^{i}+q_{3}b_{2}^{i}\rightarrow
e_{1}^{i}, $ $i=2l+1$ for $0\leq l\leq 2m$.

\textbf{Step 4. }Compute $r_{1}b_{1}^{i}+r_{3}b_{2}^{i}\rightarrow
e_{1}^{i}, $ $i=2l$ for $1\leq l\leq 2m$.

\textbf{Step 5. }Compute $q_{2}b_{1}^{i}+q_{4}b_{2}^{i}\rightarrow
e_{2}^{i}, $ $i=2l+1$ for $0\leq l\leq 2m$.

\textbf{Step 6. }Compute $r_{2}b_{1}^{i}+r_{4}b_{2}^{i}\rightarrow
e_{2}^{i}, $ $i=2l$ for $1\leq l\leq 2m$.

\textbf{Step 7.} Solve $(-1)^{n}\times
d_{i}=e_{1}^{i}(q_{2}b_{3}^{i}+q_{4}x_{i})-e_{2}^{i}(q_{1}b_{3}^{i}+q_{3}x_{i}),i=2l+1 
$ for $0\leq l\leq 2m$.

\textbf{Step 8.} Solve $5\times (-1)^{n+1}\times
d_{i}=e_{1}^{i}(r_{2}b_{3}^{i}+r_{4}x_{i})-e_{2}^{i}(r_{1}b_{3}^{i}+r_{3}x_{i}), 
$ $i=2l$ for $0\leq l\leq 2m$.

\textbf{Step 9. }Substitute for $x_{i}=b_{4}^{i}$.

\textbf{Step 10.} Construct $B_{i}$.

\textbf{Step 11.} Construct $M$.

\textbf{Step 12.} End of algorithm.

Now, we give an application of the above algorithm for $b>3$.

\begin{example}
\label{exm2} Let us consider the message matrix for the following message
text$:$%
\begin{equation*}
\text{"MIXED MODELLING FOR CRYPTOGRAPHY"}
\end{equation*}%
Using the message text, we get the following message matrix $M:$%
\begin{equation*}
M=\left[ 
\begin{array}{cccccc}
M & I & X & E & D & 0 \\ 
M & O & D & E & L & L \\ 
I & N & G & 0 & F & O \\ 
R & 0 & C & R & Y & P \\ 
T & O & G & R & A & P \\ 
H & Y & 0 & 0 & 0 & 0%
\end{array}%
\right] _{6\times 6}.
\end{equation*}%
\textbf{Coding Algorithm:}

\textbf{Step 1. }We can divide the message matrix $M$ of size $6\times 6$
into the matrices, named $B_{i}$ $\left( 1\leq i\leq 9\right) $, from left
to right, each of size is $2\times 2:$%
\begin{eqnarray*}
B_{1} &=&\left[ 
\begin{array}{cc}
M & I \\ 
M & 0%
\end{array}%
\right] \text{, }B_{2}=\left[ 
\begin{array}{cc}
X & E \\ 
D & E%
\end{array}%
\right] \text{, }B_{3}=\left[ 
\begin{array}{cc}
D & 0 \\ 
L & L%
\end{array}%
\right] , \\
B_{4} &=&\left[ 
\begin{array}{cc}
I & N \\ 
R & 0%
\end{array}%
\right] \text{, }B_{5}=\left[ 
\begin{array}{cc}
G & 0 \\ 
C & R%
\end{array}%
\right] \text{, }B_{6}=\left[ 
\begin{array}{cc}
F & O \\ 
Y & P%
\end{array}%
\right] ,\text{ } \\
B_{7} &=&\left[ 
\begin{array}{cc}
T & O \\ 
H & Y%
\end{array}%
\right] \text{, }B_{8}=\left[ 
\begin{array}{cc}
G & R \\ 
0 & 0%
\end{array}%
\right] \text{, }B_{9}=\left[ 
\begin{array}{cc}
A & P \\ 
0 & 0%
\end{array}%
\right] .\text{ }
\end{eqnarray*}

\textbf{Step 2.} Due to $b=9>3$, we calculate $n=\left[ \left\vert \frac{b}{2%
}\right\vert \right] =4$. For $n=4$, we use the following \textquotedblleft
character table\textquotedblright\ for the message matrix $M:$%
\begin{equation*}
\begin{tabular}{|l|l|l|l|l|l|l|l|l|l|l|l|}
\hline
$M$ & $I$ & $X$ & $E$ & $D$ & $0$ & $M$ & $O$ & $D$ & $E$ & $L$ & $L$ \\ 
\hline
$16$ & $12$ & $27$ & $8$ & $7$ & $0$ & $16$ & $18$ & $7$ & $8$ & $15$ & $15$
\\ \hline
$I$ & $N$ & $G$ & $0$ & $F$ & $O$ & $R$ & $0$ & $C$ & $R$ & $Y$ & $P$ \\ 
\hline
$12$ & $17$ & $10$ & $0$ & $9$ & $18$ & $21$ & $0$ & $6$ & $21$ & $28$ & $19$
\\ \hline
$T$ & $O$ & $G$ & $R$ & $A$ & $P$ & $H$ & $Y$ & $0$ & $0$ & $0$ & $0$ \\ 
\hline
$23$ & $18$ & $10$ & $21$ & $4$ & $19$ & $11$ & $28$ & $0$ & $0$ & $0$ & $0$
\\ \hline
\end{tabular}%
.
\end{equation*}

\textbf{Step 3.} We have the elements of the blocks $B_{i}$ $\left( 1\leq
i\leq 9\right) $ as follows$:$%
\begin{equation*}
\begin{tabular}{|l|l|l|l|}
\hline
$b_{1}^{1}=16$ & $b_{2}^{1}=12$ & $b_{3}^{1}=16$ & $b_{4}^{1}=18$ \\ \hline
$b_{1}^{2}=27$ & $b_{2}^{2}=8$ & $b_{3}^{2}=7$ & $b_{4}^{2}=8$ \\ \hline
$b_{1}^{3}=7$ & $b_{2}^{3}=0$ & $b_{3}^{3}=15$ & $b_{4}^{3}=15$ \\ \hline
$b_{1}^{4}=12$ & $b_{2}^{4}=17$ & $b_{3}^{4}=21$ & $b_{4}^{4}=0$ \\ \hline
$b_{1}^{5}=10$ & $b_{2}^{5}=0$ & $b_{3}^{5}=6$ & $b_{4}^{5}=21$ \\ \hline
$b_{1}^{6}=9$ & $b_{2}^{6}=18$ & $b_{3}^{6}=28$ & $b_{4}^{6}=19$ \\ \hline
$b_{1}^{7}=23$ & $b_{2}^{7}=18$ & $b_{3}^{7}=11$ & $b_{4}^{7}=28$ \\ \hline
$b_{1}^{8}=10$ & $b_{2}^{8}=21$ & $b_{3}^{8}=0$ & $b_{4}^{8}=0$ \\ \hline
$b_{1}^{9}=4$ & $b_{2}^{9}=19$ & $b_{3}^{9}=0$ & $b_{4}^{9}=0$ \\ \hline
\end{tabular}%
.
\end{equation*}

\textbf{Step 4.} Now we calculate the determinants $d_{i}$ of the blocks $%
B_{i}:$%
\begin{equation*}
\begin{tabular}{|l|}
\hline
$d_{1}=\det (B_{1})=96$ \\ \hline
$d_{2}=\det (B_{2})=160$ \\ \hline
$d_{3}=\det (B_{3})=105$ \\ \hline
$d_{4}=\det (B_{4})=-357$ \\ \hline
$d_{5}=\det (B_{5})=210$ \\ \hline
$d_{6}=\det (B_{6})=-333$ \\ \hline
$d_{7}=\det (B_{7})=446$ \\ \hline
$d_{8}=\det (B_{8})=0$ \\ \hline
$d_{9}=\det (B_{9})=0$ \\ \hline
\end{tabular}%
.
\end{equation*}

\textbf{Step 5.} Using Step 3 and Step 4 we obtain the following matrix $F:$%
\begin{equation*}
F=\left[ 
\begin{array}{cccc}
96 & 16 & 12 & 16 \\ 
160 & 27 & 8 & 7 \\ 
105 & 7 & 0 & 15 \\ 
-357 & 12 & 17 & 21 \\ 
210 & 10 & 0 & 6 \\ 
-333 & 9 & 18 & 28 \\ 
446 & 23 & 18 & 11 \\ 
0 & 10 & 21 & 0 \\ 
0 & 4 & 19 & 0%
\end{array}%
\right] .
\end{equation*}

\textbf{Step 6.} End of algorithm.

\textbf{Decoding algorithm:}

\textbf{Step 1.} It is known that%
\begin{equation*}
Q^{4}=\left[ 
\begin{array}{cc}
F_{5} & F_{4} \\ 
F_{4} & F_{3}%
\end{array}%
\right] =\left[ 
\begin{array}{cc}
5 & 3 \\ 
3 & 2%
\end{array}%
\right] \text{.}
\end{equation*}

\textbf{Step 2. }It is known that%
\begin{equation*}
R_{4}=RQ^{4}=\left[ 
\begin{array}{cc}
L_{5} & L_{4} \\ 
L_{4} & L_{3}%
\end{array}%
\right] =\left[ 
\begin{array}{cc}
1 & 2 \\ 
2 & -1%
\end{array}%
\right] \left[ 
\begin{array}{cc}
5 & 3 \\ 
3 & 2%
\end{array}%
\right] =\left[ 
\begin{array}{cc}
11 & 7 \\ 
7 & 4%
\end{array}%
\right] \text{.}
\end{equation*}

\textbf{Step 3.} If $i$ is an odd number, we use the Fibonacci $Q$-matrix.
Now we compute the elements $e_{1}^{i},$ for $i=1,3,5,7,9,$ in order to
construct the matrix $E_{i}:$%
\begin{equation*}
e_{1}^{1}=116\text{, }e_{1}^{3}=35\text{, }e_{1}^{5}=50,\text{ }e_{1}^{7}=169%
\text{ and }e_{1}^{9}=47\text{.}
\end{equation*}

\textbf{Step 4}. If $i$ is an even number, we use the $R$-matrix. Now we
compute the elements $e_{1}^{i},$ for $i=2,4,6,8$ in order to construct the
matrix $E_{i}:$%
\begin{equation*}
e_{1}^{2}=353\text{, }e_{1}^{4}=251\text{, }e_{1}^{6}=225\text{ and }%
e_{1}^{8}=257\text{.}
\end{equation*}

\textbf{Step 5. }If $i$ is an odd number, we use the Fibonacci $Q$-matrix.
Now we compute the elements $e_{2}^{i},$ for $i=1,3,5,7,9,$ in order to
construct the matrix $E_{i}:$%
\begin{equation*}
e_{2}^{1}=72,e_{2}^{3}=21,e_{2}^{5}=30,e_{2}^{7}=105\text{ and }e_{2}^{9}=30%
\text{.}
\end{equation*}

\textbf{Step 6. }If $i$ is an even number, we use the $R$-matrix. Now we
compute the elements $e_{2}^{i},$ for $i=2,4,6,8,$ in order to construct the
matrix $E_{i}:$%
\begin{equation*}
e_{2}^{2}=221\text{, }e_{2}^{4}=152\text{, }e_{2}^{6}=135,\text{ and }%
e_{2}^{8}=154\text{.}
\end{equation*}

\textbf{Step 7.} If $i$ is an odd number, we use the Fibonacci $Q$-matrix.
Now, we calculate the elements $x_{i}$ for $i=1,3,5,7,9.$%
\begin{eqnarray*}
(-1)^{4}96 &=&116(48+2x_{1})-72(80+3x_{1}) \\
&\Rightarrow &x_{1}=18\text{.}
\end{eqnarray*}%
\begin{eqnarray*}
(-1)^{4}105 &=&35(45+2x_{3})-21(75+3x_{3}) \\
&\Rightarrow &x_{3}=15\text{.}
\end{eqnarray*}%
\begin{eqnarray*}
(-1)^{4}210 &=&50(18+2x_{5})-30(30+3x_{5}) \\
&\Rightarrow &x_{5}=21\text{.}
\end{eqnarray*}%
\begin{eqnarray*}
(-1)^{4}446 &=&169(33+2x_{7})-105(55+3x_{7}) \\
&\Rightarrow &x_{7}=28\text{.}
\end{eqnarray*}%
\begin{eqnarray*}
(-1)^{4}0 &=&47(0+2x_{9})-30(0+3x_{9}) \\
&\Rightarrow &x_{9}=0\text{.}
\end{eqnarray*}

\textbf{Step 8.} If $i$ is an even number, we use the $R-$matrix. Now, we
calculate the elements $x_{i}$ for $i=2,4,6,8.$%
\begin{eqnarray*}
5(-1)^{5}160 &=&353(49+4x_{2})-221(77+7x_{2}) \\
&\Rightarrow &x_{2}=8\text{.}
\end{eqnarray*}%
\begin{eqnarray*}
5(-1)^{5}\left( -357\right) &=&251(147+4x_{4})-152(231+3x_{4}) \\
&\Rightarrow &x_{4}=0\text{.}
\end{eqnarray*}%
\begin{eqnarray*}
5(-1)^{5}\left( -333\right) &=&225(196+4x_{6})-135(308+7x_{6}) \\
&\Rightarrow &x_{6}=19\text{.}
\end{eqnarray*}%
\begin{eqnarray*}
5(-1)^{5}0 &=&257(0+4x_{8})-154(0+7x_{8}) \\
&\Rightarrow &x_{8}=0\text{.}
\end{eqnarray*}

\textbf{Step 9.} We rename $x_{i}$ as follows$:$%
\begin{eqnarray*}
x_{1} &=&b_{4}^{1}=18\text{, }x_{2}=b_{4}^{2}=8\text{, }x_{3}=b_{4}^{3}=15,%
\text{ }x_{4}=b_{4}^{4}=0\text{, }x_{5}=b_{4}^{5}=21, \\
x_{6} &=&b_{4}^{6}=19\text{, }x_{7}=b_{4}^{7}=28\text{, }x_{8}=b_{4}^{8}=0%
\text{ and }x_{9}=b_{4}^{9}=0\text{. }
\end{eqnarray*}

\textbf{Step 10. }We construct the block matrices $B_{i}:$%
\begin{eqnarray*}
B_{1} &=&\left[ 
\begin{array}{cc}
16 & 12 \\ 
16 & 18%
\end{array}%
\right] \text{, }B_{2}=\left[ 
\begin{array}{cc}
27 & 8 \\ 
7 & 8%
\end{array}%
\right] \text{, }B_{3}=\left[ 
\begin{array}{cc}
7 & 0 \\ 
15 & 15%
\end{array}%
\right] ,\text{ } \\
B_{4} &=&\left[ 
\begin{array}{cc}
12 & 17 \\ 
4 & 0%
\end{array}%
\right] \text{, }B_{5}=\left[ 
\begin{array}{cc}
10 & 0 \\ 
6 & 21%
\end{array}%
\right] \text{, }B_{6}=\left[ 
\begin{array}{cc}
9 & 18 \\ 
28 & 19%
\end{array}%
\right] ,\text{ } \\
B_{7} &=&\left[ 
\begin{array}{cc}
23 & 18 \\ 
11 & 28%
\end{array}%
\right] \text{, }B_{8}=\left[ 
\begin{array}{cc}
10 & 21 \\ 
0 & 0%
\end{array}%
\right] \text{, }B_{9}=\left[ 
\begin{array}{cc}
4 & 9 \\ 
0 & 0%
\end{array}%
\right] .\text{ }
\end{eqnarray*}

\textbf{Step 11.} We obtain the following message matrix $M:$%
\begin{equation*}
M=\left[ 
\begin{array}{cccccc}
16 & 12 & 27 & 8 & 7 & 0 \\ 
16 & 18 & 7 & 8 & 15 & 15 \\ 
12 & 17 & 10 & 0 & 9 & 18 \\ 
4 & 0 & 6 & 21 & 28 & 19 \\ 
23 & 18 & 10 & 21 & 4 & 9 \\ 
11 & 28 & 0 & 0 & 0 & 0%
\end{array}%
\right] =\left[ 
\begin{array}{cccccc}
M & I & X & E & D & 0 \\ 
M & O & D & E & L & L \\ 
I & N & G & 0 & F & O \\ 
R & 0 & C & R & Y & P \\ 
T & O & G & R & A & P \\ 
H & Y & 0 & 0 & 0 & 0%
\end{array}%
\right] .
\end{equation*}

\textbf{Step 9.} End of algorithm.
\end{example}

\section{Comparisons and Conclusion}

\label{sec:3} In this section we give the differences between the method
given in \cite{Tas} and the above methods. At first, in \cite{Tas}, the
number $n$ is defined by 
\begin{equation*}
n=\left\{ 
\begin{array}{ccc}
3 & \text{,} & b\leq 3 \\ 
b & \text{,} & b>3%
\end{array}%
\right. \text{.}
\end{equation*}%
On the other hand, in our methods the number $n$ has been given in different
ways as we have explained in Section \ref{sec:1}. Because of the selection
method of $n,$ we are studying more smaller numbers to calculate the
matrices $Q^{n}$, $R_{n}$ and to form the character table. Hence we obtain
more easier methods than the method given in \cite{Tas}. Furthermore if we
use the minesweeper model, then the security has been increased according to
Lucas Blocking Model given in Section \ref{sec:1} and Fibonacci Blocking
Model given in \cite{Tas}. Especially to increase the security it can be
changed the decoding method of blocks using Fibonacci and Lucas numbers
randomly.

\end{document}